# Variational Calculation of Effective Classical Potential at $T \neq 0$ to Higher Orders

H.Kleinert          H.Meyer

November 12, 1993

**Abstract**

Using the new variational approach proposed recently for a systematic improvement of the locally harmonic Feynman–Kleinert approximation to path integrals we calculate the partition function of the anharmonic oscillator for all temperatures and coupling strength with high accuracy.

1) Some time ago, Feynman and Kleinert [1] extended the variational approach to euclidean path integrals developed earlier by Feynman in his textbook on statistical mechanics [2], improving greatly the accuracy at low temperatures. A similar extension was found independently by Giacetti and Tognetti [3] who applied this method to several statistical systems [4]. The approach is particularly successful in systems in which the quantum effects do not produce essentially new phenomena, their main result being a modification of the quasi-harmonic properties of a system. This happens in quantum crystals [5] where experimental data can now be explained very well. The approach yields a good approximation to the *effective classical potential* $V_{\text{eff,cl}}$ at all temperatures and serves to calculate the free energy of the system as well as particle distributions at all coupling strength, including the strong coupling limit.

The purpose of this note is to present results of a recently proposed systematic improvement [6] of this method for the anharmonic oscillator carried out to third order in the coupling strength. The results turn out to be in excellent agreement with a precise numerical values of the free energy [7] at all temperatures and coupling strengths available in the literature.

2) Let us briefly review the Feynman–Kleinert approach. The aim is to write the partition function as a classical configuration space integral over a Boltzmann factor involving an effective classical potential $V_{\text{eff,cl}}(x_0)$:

$$Z = \int \frac{dx_0}{\sqrt{2\pi\hbar^2\beta/M}} e^{-\beta V_{\text{eff,cl}}(x_0)} \tag{1}$$

with $\beta \equiv 1/k_B T$. The variable $x_0$ is equal to the time-averaged position $\bar{x} = \int_0^{\hbar\beta} d\tau\, x(\tau)$ of the fluctuating path. The effective potential $V_{\text{eff,cl}}(x_0)$ is defined by the path integral

$$\exp\left\{-\beta V_{\text{eff,cl}}\right\} = Z^{x_0} = \oiint \mathcal{D}x\, \bar{\delta}(\bar{x} - x_0) \exp\left\{-\mathcal{A}/\hbar\right\}, \qquad (2)$$

with the euclidean action

$$\mathcal{A} = \int_0^{\hbar\beta} d\tau \left\{\frac{1}{2}M\dot{x}^2 + V(x)\right\}$$

where

$$\bar{\delta}(\bar{x} - x_0) = \sqrt{2\pi\hbar^2\beta/M}\, \delta(\bar{x} - x_0)$$

restricts the path average $\bar{x}$ to the value $x_0$. The paths are periodic in $\hbar\beta$. We shall refer to $Z^{x_0}$ as a *restricted partition function*. The usefulness of separating out $\bar{x}$ derives from the fact that at any temperature the fluctuations rarely carry $x(\tau)$ far away from $\bar{x}$. This allows approximating the deviations from $\bar{x}$ in a quasi-harmonic way. One reexpresses the local partition function as an expectation value within a harmonic trial system whose action is, for each $x_0$,

$$\mathcal{A}_\Omega^{x_0} = \frac{M}{2} \int_0^{\hbar\beta} d\tau \left\{\dot{x}^2 + \Omega^2(x_0)(x - x_0)^2\right\} \qquad (3)$$

and chooses $\Omega^2(x_0)$ optimally. With the help of $\mathcal{A}_\Omega^{x_0}$, the defining expression (2) can be rewritten as

$$\exp\left\{-\beta V_{\text{eff,cl}}(x_0)\right\} = (Z_\Omega^{x_0})^{-1} \left\langle \exp\left\{-(\mathcal{A} - \mathcal{A}_\Omega^{x_0})/\hbar\right\}\right\rangle_\Omega^{x_0} \qquad (4)$$

for any $\Omega(x_0)$. The subscript $\Omega$ stands for the trial system in which the expectation is calculated and the superscript $x_0$ indicates the restricted value of $\bar{x}$.

The expectation value on the right-hand side of (4) cannot be calculated exactly. A lowest-order approximation is found with the help of the Jensen-Peierls inequality

$$\left\langle \exp\left\{-(\mathcal{A} - \mathcal{A}_\Omega^{x_0})/\hbar\right\}\right\rangle_\Omega^{x_0} \geq \exp\left\{-\left\langle \mathcal{A} - \mathcal{A}_\Omega^{x_0}\right\rangle_\Omega^{x_0}/\hbar\right\}.$$

The harmonic expectation value in the exponent of the right-hand side yields the approximation $W_1$ for the effective classical potential

$$V_{\text{eff,cl}}(x_0) \approx W_1(x_0) = V_\Omega^{x_0}(x_0) + V_{a^2}(x_0) - \frac{1}{2}M\Omega^2(x_0)a^2(x_0). \qquad (5)$$

The first term is the logarithm of the restricted partition function of the trial system $-\ln\left\{(\hbar\beta\Omega(x_0)/2)/\sinh(\hbar\beta\Omega(x_0)/2)\right\}/\beta$, the last two terms are the restricted expectations

$$\langle V(x)\rangle_\Omega^{x_0} = V_{a^2}(x_0)$$

and

$$\frac{1}{2}M\Omega^2(x_0)\left\langle (x-x_0)^2\right\rangle_\Omega^{x_0} = \frac{1}{2}M\Omega^2(x_0)a^2(x_0).$$

In the last equation we have abbreviated the restricted square deviation $\left\langle (x-x_0)^2\right\rangle_\Omega^{x_0}$ by

$$a^2(x_0) = \frac{1}{M\beta\Omega^2}\left\{\frac{\hbar\beta\Omega}{2}\coth\frac{\hbar\beta\Omega}{2} - 1\right\}. \tag{6}$$

The first term is the usual thermal expectation $<x^2(\tau)>$; the second term subtracts from this the square deviations of the temporal average $\bar{x}$ from $x_0$. The restricted expectation of the potential is obtained by a simple gaussian smearing process of a width $a^2(x_0)$

$$V_{a^2}(x_0) = \int_{-\infty}^{\infty}\frac{dx}{\sqrt{2\pi a^2}}V(x)e^{-(x-x_0)^2/2a^2}. \tag{7}$$

The best approximation is obtained by minimizing the function $W_1(x_0)$ with respect to $\Omega(x_0)$ which yields the condition

$$\Omega^2 = \frac{2}{M}\frac{\partial}{\partial a^2}V_{a^2}(x_0). \tag{8}$$

The resulting $W_1(x_0)$ is always slightly larger than the exact effective classical potential $V_{\text{eff,cl}}(x_0)$.

Eqs. (6) and (8) are solved numerically. The resulting approximation to the partition function

$$Z_1 = \int_{-\infty}^{\infty}\frac{dx_0}{\sqrt{2\pi\hbar^2\beta/M}}e^{-\beta W_1(x_0)}$$

leads to the free energy $F_1 = -k_\text{B}T\ln Z_1$ which describes the true free energy $F = -k_\text{B}T\ln Z$ of the system quite well at all temperatures. Ref. [8] explains why this is so.

For the anharmonic oscillator potential

$$V(x) = \frac{\omega^2}{2}x^2 + \frac{g}{4}x^4,$$

Eq. (8) leads to the following equation for the optimal $\Omega$:

$$\Omega^2 = 3gx_0^2 + 3g\frac{1}{\beta\Omega^2}\left\{\frac{\hbar\beta\Omega}{2}\coth\frac{\hbar\beta\Omega}{2} - 1\right\} + 1. \tag{9}$$

The resulting free energies are listed in Tab. 1 for comparison with the improved results to be derived in this note.

3) To go beyond the above approximation we split the action into a free part $\mathcal{A}_0$ and an interacting part $\mathcal{A}_{\text{int}}^{x_0}$. The first contains all classical terms depending on $x_0$ plus the trial action $\mathcal{A}_\Omega^{x_0}$ of Eq. (3), the second term all the rest. The partition function can then be rewritten as

$$Z = \int_{-\infty}^{\infty} \frac{dx_0}{\sqrt{2\pi\hbar^2\beta/M}} Z^{x_0} = \int_{-\infty}^{\infty} \frac{dx_0}{\sqrt{2\pi\hbar^2\beta/M}} e^{V(x_0)/\hbar} \left\langle e^{-\mathcal{A}_{\text{int}}^{x_0}/\hbar} \right\rangle_\Omega^{x_0}. \quad (10)$$

In the previous section, the Jensen–Peierls inequality was used to approximate the expectation value on the right-hand side of (10). Now we evaluate this expectation perturbatively, expanding the exponential function in a Taylor series

$$Z^{x_0} = e^{-\beta V(x_0)} Z_\Omega^{x_0} \left\{ 1 - \frac{1}{\hbar} \langle \mathcal{A}_{\text{int}}^{x_0} \rangle_\Omega^{x_0} + \frac{1}{2!\hbar^2} \langle (\mathcal{A}_{\text{int}}^{x_0})^2 \rangle_\Omega^{x_0} - \frac{1}{3!\hbar^3} \langle (\mathcal{A}_{\text{int}}^{x_0})^3 \rangle_\Omega^{x_0} + \cdots \right\}.$$

and going over to the connected parts by a reexpansion into cumulants

$$\begin{aligned}Z^{x_0} = \ &\exp\Big\{ -\beta V(x_0) - \beta V_\Omega^{x_0} - \frac{1}{\hbar} \langle \mathcal{A}_{\text{int}}^{x_0} \rangle_\Omega^{x_0} + \frac{1}{2\hbar^2} \langle (\mathcal{A}_{\text{int}}^{x_0})^2 \rangle_{\Omega,c}^{x_0} \\ & - \frac{1}{6\hbar^3} \langle (\mathcal{A}_{\text{int}}^{x_0})^3 \rangle_{\Omega,c}^{x_0} + \cdots \Big\}. \end{aligned} \quad (11)$$

The truncated exponents

$$\begin{aligned}W_N(x_0) = \ & V(x_0) + V_\Omega^{x_0} + \frac{1}{\hbar\beta} \langle \mathcal{A}_{\text{int}}^{x_0} \rangle_\Omega^{x_0} - \frac{1}{2\hbar^2\beta} \langle (\mathcal{A}_{\text{int}}^{x_0})^2 \rangle_{\Omega,c}^{x_0} + \frac{1}{6\hbar^3\beta} \langle (\mathcal{A}_{\text{int}}^{x_0})^3 \rangle_{\Omega,c}^{x_0} \\ & \cdots + \frac{(-1)^N}{N!\hbar^N\beta} \langle (\mathcal{A}_{\text{int}}^{x_0})^N \rangle_\Omega^{x_0} \end{aligned} \quad (12)$$

are optimized in $\Omega(x_0)$ and yield the successive approximations to the effective classical potential $V_{\text{eff,cl}}(x_0)$. The lowest approximation $W_1(x_0)$ obviously coincides with the Feynman-Kleinert approximation.

We study the behavior of $W_N(x_0)$ up to $N = 3$. The condition of an optimal $\Omega(x_0)$ cannot always be met by an extremum of $W_N(x_0)$, since for higher $N > 1$ the condition $\partial W_N(x_0)/\partial\Omega(x_0) = 0$ does not always have real solutions $\Omega(x_0)$. This happens for example at $N = 2$. In such cases, an optimal result is obtained by enforcing a minimal dependence on $\Omega$ via the disappearance of the second derivative of $W_N(x_0)$ with respect to $\Omega(x_0)$.

To calculate the restricted expectation values in equation (12) we consider polynomial-like potentials in view of the intended quantum field theoretic applications. The interaction $\mathcal{A}_{\text{int}}^{x_0}$ is expanded around $x_0$

$$\mathcal{A}_{\text{int}}^{x_0} = \int_0^{\hbar\beta} d\tau \left\{ \frac{g_2}{2!}(x - x_0)^2 + \frac{g_3}{3!}(x - x_0)^3 + \frac{g_4}{4!}(x - x_0)^4 \cdots \right\}, \quad (13)$$

defining the coupling constants

$$g_i = V^{(i)}(x_0) - \Omega^2 \delta_{i,2}. \quad (14)$$

Then (12) is obtained by calculating all connected vacuum diagrams formed with vertices $g_n/n!\hbar$ and lines which stand for the correlation function

$$G^{(2)}(\tau,\tau') = \frac{1}{M\beta\Omega^2}\left\{\frac{\hbar\beta\Omega}{2}\frac{\cosh[\frac{\hbar\beta\Omega}{2}(1-\frac{2|\tau-\tau'|}{\hbar\beta})]}{\sinh\frac{\hbar\beta\Omega}{2}} - 1\right\}, \qquad (15)$$

in which the first term is the usual Green function $\langle x(\tau)x(\tau')\rangle$ of the harmonic oscillator while the second term subtracts from this the zero-frequency part $\langle x_0^2\rangle = 1/\beta\Omega^2$.

The number of Feynman integrals to be calculated is reduced in three ways. First, only one-particle irreducible vacuum graphs contribute (i.e., there are no tadpole diagrams) due to the absence of the zero-frequency modes in $\delta x(\tau) = x(\tau) - x_0$. Second, Feynman integrals of subdiagrams which touch the rest in one vertex can be factored out. Third, diagrams with vertices $g_2$ (mass insertions) can be obtained from those without these vertices by replacing the frequency $\Omega$ by $\tilde{\Omega} = \sqrt{\Omega^2 + g_2}$ and expanding the result in powers of $g_2$.

4) We now turn to the application to an anharmonic oscillator with the potential

$$V(x) = \frac{\omega^2}{2}x^2 + \frac{g}{4}x^4.$$

The interaction is

$$\mathcal{A}_{\text{int}}^{x_0} = \int d\tau \left\{\frac{g_2}{2!}(\delta x)^2(\tau) + \frac{g_3}{3!}(\delta x)^3(\tau) + \frac{g_4}{4!}(\delta x)^4(\tau)\right\}$$

(we use natural units with $\hbar = M = 1$), with the coupling constants

$$\begin{aligned} g_2 &= (\omega^2 - \Omega^2) + 3gx_0^2 \\ g_3 &= 6gx_0 \\ g_4 &= 6g. \end{aligned} \qquad (16)$$

The quadratic coupling $g_2(\delta x)^2(\tau)/2$ is now included into the harmonic trial potential term by changing it to $\tilde{\Omega}^2(x-x_0)^2/2$ with the modified frequency $\tilde{\Omega}$ introduced above. The third-order approximation

$$W_3(\tilde{\Omega}) = V(x_0) + V_{\tilde{\Omega}}^{x_0} + \frac{1}{\hbar\beta}\langle\mathcal{A}_{\text{int}}^{x_0}\rangle_{\tilde{\Omega}}^{x_0} - \frac{1}{2\hbar^2\beta}\langle(\mathcal{A}_{\text{int}}^{x_0})^2\rangle_{\tilde{\Omega},c}^{x_0} + \frac{1}{6\hbar^3\beta}\langle(\mathcal{A}_{\text{int}}^{x_0})^3\rangle_{\tilde{\Omega},c}^{x_0} \qquad (17)$$

has the diagrammatic expansion

$$W_3(\tilde{\Omega}) = V(x_0) + \frac{1}{\beta}\left\{-\frac{1}{2}\bigcirc^{\tilde{\Omega}} + \underset{3}{\infty}\right.$$

$$\left. -\frac{1}{2!}\left(\underset{6}{\ominus}^{\tilde{\Omega}} + \underset{24}{\ominus}^{\tilde{\Omega}} + \underset{72}{\infty}^{\tilde{\Omega}}\right)\right.$$

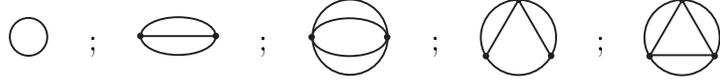

The number under each diagram count in how many Wick contractions it appears; this number multiplies the prefactors.

Only five integrals corresponding to the diagrams need to be evaluated explicitly. The remaining diagrams follow by differentiation with respect to $\Omega^2$ and factorization into subdiagrams touching each other at a point.

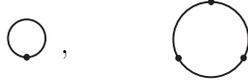

In quantum field theory one is used to calculate Feynman diagrams in momentum space. At a finite temperature this requires performing multiple sums over Matsubara frequencies. To higher order, they are hard to do in closed form. In the present case of a $D = 1$ -dimensional quantum field theory it is more convenient to evaluate the diagrams in $\tau$-space as integrals over products of the Green function (15). For example, the diagrams require the integrals (before replacing $\Omega \to \tilde{\Omega}$).

$$\hbar \beta a^2 \equiv \int_0^{\hbar\beta} G^{(2)}(\tau,\tau)^2 d\tau,$$

$$\hbar\beta \left(\frac{1}{\Omega}\right)^2 a_3^6 \equiv \int_0^{\hbar\beta}\int_0^{\hbar\beta}\int_0^{\hbar\beta} G^{(2)}(\tau_1,\tau_2) G^{(2)}(\tau_2,\tau_3) G^{(2)}(\tau_3,\tau_1)\, d\tau_1 d\tau_2 d\tau_3,$$

respectively. The factors $\hbar\beta$ are due to an overall $\tau$-integral, the powers of $1/\Omega$ arise from the remaining ones. The parameters $a_m$ have the dimension of a length, the subscript indicating the number of vertices in the diagram (defining $a_1^2 \equiv a^2$). In terms of these two Feynman integrals, the integral associated with the last diagram in (18) is given by the product

$$\hat{=} \;\; \hbar\beta \left(\frac{1}{\Omega}\right)^2 a_3^6 (a^2)^3$$

The symbol $\hat{=}$ indicates that the right-hand side is only the Feynman integral of the diagram without the factors $g_n/n!\hbar$ associated with the vertices.

The explicit expressions for the different Feynman integrals are, with the abbreviation $x = \hbar\beta\Omega$,

$$\bigcirc \;=\; \text{Tr}\log G^{(2)} = -2\beta V_\Omega^{x_0} = -2\log\frac{\sinh(x/2)}{x/2}$$

$$\ominus \;\hat{=}\; \hbar\beta\left(\frac{1}{\Omega}\right)a_2^6$$

$$a_2^6 \;=\; \left(\frac{\hbar}{\Omega}\right)^3 \frac{1}{24x^2}\frac{1}{\sinh^2\frac{x}{2}}\left\{-24 - 4\,x^2 + 24\,\cosh x + x^2\cosh x - 9\,x\sinh x\right\}$$

$$\ominus \;\hat{=}\; \hbar\beta\left(\frac{1}{\Omega}\right)a_2^8$$

$$a_2^8 \;=\; \left(\frac{\hbar}{\Omega}\right)^4 \frac{1}{768x^3}\frac{1}{\sinh^4\frac{x}{2}}\left\{-864 + 18\,x^4 + 1152\,\cosh x + 32\,x^2\cosh x\right.$$
$$-288\,\cosh 2x - 32\,x^2\cosh 2x - 288\,x\sinh x + 24\,x^3\sinh x$$
$$\left.+144\,x\sinh 2x + 3\,x^3\sinh 2x\right\}$$

$$\triangle \;\hat{=}\; \hbar\beta\left(\frac{1}{\Omega}\right)^2 a_3^{10}$$

$$a_3^{10} \;=\; \left(\frac{\hbar}{\Omega}\right)^5 \frac{1}{2304x^3}\frac{1}{\sinh^4\frac{x}{2}}\left\{-3456 - 414\,x^2 - 6\,x^4 + 4608\,\cosh x\right.$$
$$+496\,x^2\cosh x - 1152\,\cosh 2x - 82\,x^2\cosh 2x - 1008\,x\sinh x$$
$$\left.-16\,x^3\sinh x + 504\,x\sinh 2x + 5\,x^3\sinh 2x\right\}$$

$$\triangle \;\hat{=}\; \hbar\beta\left(\frac{1}{\Omega}\right)^2 a_3^{12}$$

$$a_3^{12} \;=\; \left(\frac{\hbar}{\Omega}\right)^6 \frac{1}{49152x^4}\frac{1}{\sinh^6\frac{x}{2}}\left\{-107520 - 7360\,x^2 + 624\,x^4 + 96\,x^6\right.$$
$$+161280\,\cosh x + 12000\,x^2\cosh x - 777\,x^4\cosh x + 24\,x^6\cosh x$$
$$-64512\,\cosh 2x - 5952\,x^2\cosh 2x + 144\,x^4\cosh 2x + 10752\,\cosh 3x$$
$$+1312\,x^2\cosh 3x + 9\,x^4\cosh 3x - 28800\,x\sinh x + 1120\,x^3\sinh x$$
$$+324\,x^5\sinh x\, 23040\,x\sinh 2x - 320\,x^3\sinh 2x - 5760\,x\sinh 3x$$
$$\left.-160\,x^3\sinh 3x\right\} \tag{18}$$

To obtain the diagrams involving $g_2$ we replace, as explained above, $\Omega$ by $\tilde{\Omega}$ and expand in powers of $g_2/2\hbar$:

$$\underset{3}{\overset{\tilde{\Omega}}{\infty}} = \underset{3}{\infty} - \underset{24}{\infty} + \frac{1}{6}\underset{72}{\infty} + \frac{1}{6}\underset{144}{\infty} ,$$

$$\frac{1}{2}\underset{6}{\overset{\tilde{\Omega}}{\bigodot}} = \frac{1}{2}\underset{6}{\bigodot} - \frac{1}{6}\underset{108}{\bigodot} ,$$

$$\frac{1}{2}\underset{24}{\overset{\tilde{\Omega}}{\bigodot}} = \frac{1}{2}\underset{24}{\bigodot} - \frac{1}{6}\underset{576}{\bigodot} ,$$

$$\frac{1}{2}\underset{72}{\overset{\tilde{\Omega}}{\infty\!\infty}} = \frac{1}{2}\underset{72}{\infty\!\infty} - \frac{1}{6}\underset{864}{\infty\!\infty} - \frac{1}{6}\underset{864}{\infty\!\infty} .$$

Note that the dot being associated with the vertex $g_2/2$ is equal to *half* a mass insertion. The numbers underneath the diagrams are their multiplicities; they act as factors.

The expansion of the trace log follows a different pattern since the one-loop diagram requires a reexpansion of the logarithm pictured by it:

$$\overset{\tilde{\Omega}}{\bigcirc} = \bigcirc - 2\,\bigcirc + \underset{2}{\bigcirc} - \frac{1}{3}\underset{8}{\bigcirc} .$$

As an example, consider the Feynman diagram

$$\underset{2}{\bigcirc} = 2\left(\frac{g_2}{2\hbar}\right)^2 \hbar\beta\frac{1}{\Omega}a_2^4.$$

It is obtained from the second-order Taylor expansion term of the trace log as follows:

$$2\left(\frac{g_2}{2\hbar}\right)^2 \hbar\beta\frac{1}{\Omega}a_2^4 = \frac{1}{2!}\frac{\partial^2}{(\partial\tilde{\Omega}^2)^2}\left[-2\beta V_{\tilde{\Omega}}^{x_0}\right]_{\tilde{\Omega}^2=\Omega^2} (\tilde{\Omega}^2 - \Omega^2)^2.$$

The differentiation yields

$$a_2^4 = \left(\frac{\hbar}{\Omega}\right)^2 \frac{1}{8x}\frac{1}{\sinh^2\frac{x}{2}}\left\{4 + x^2 - 4\,\cosh x + x\sinh x\right\},$$

corresponding to the Feynman integral

$$\hbar\beta\frac{1}{\Omega}a_2^4 = \int_0^{\hbar\beta}\int_0^{\hbar\beta} G^2(\tau_2 - \tau_1)\,d\tau_1 d\tau_2.$$

Similarly we find the Feynman integrals

$$\bigcirc \;\hat{=}\; \hbar\beta\left(\frac{1}{\Omega}\right) a_3^6$$

$$a_3^6 = \left(\frac{\hbar}{\Omega}\right)^3 \frac{1}{64x}\frac{1}{\sinh^3\frac{x}{2}}\left\{-3\,x\cosh\frac{x}{2} + 2\,x^3\cosh\frac{x}{2} + 3\,x\cosh\frac{3x}{2}\right. \quad (19)$$

$$\left. -48\,\sinh\frac{x}{2} + 6\,x^2\sinh\frac{x}{2} - 16\,\sinh\frac{3x}{2}\right\}$$

$$\ominus \;\hat{=}\; \hbar\beta\left(\frac{1}{\Omega}\right)^2 a_3^8$$

$$a_3^8 = \left(\frac{\hbar}{\Omega}\right)^4 \frac{1}{288x^2}\frac{1}{\sinh^3\frac{x}{2}}\left\{45\,x\cosh\frac{x}{2} - 6\,x^3\cosh\frac{x}{2} - 45\,x\cosh\frac{3x}{2}\right. \quad (20)$$

$$\left. -432\,\sinh\frac{x}{2} - 54\,x^2\sinh\frac{x}{2} + 144\,\sinh\frac{3x}{2} + 4\,x^2\sinh\frac{3x}{2}\right\}$$

$$\ominus\!\!\!- \;\hat{=}\; \hbar\beta\left(\frac{1}{\Omega}\right)^2 a_{3'}^{10}$$

$$a_{3'}^{10} = \left(\frac{\hbar}{\Omega}\right)^5 \frac{1}{4096x^3}\frac{1}{\sinh^5\frac{x}{2}}\left\{672\,x\cosh\frac{x}{2} - 8\,x^3\cosh\frac{x}{2} + 24\,x^5\cosh\frac{x}{2}\right.$$

$$-1008\,x\cosh\frac{3x}{2} + 3\,x^3\cosh\frac{3x}{2} + 336\,x\cosh\frac{5x}{2} + 5\,x^3\cosh\frac{5x}{2}$$

$$-7680\,\sinh\frac{x}{2} - 352\,x^2\sinh\frac{x}{2} + 72\,x^4\sinh\frac{x}{2} + 3840\,\sinh\frac{3x}{2}$$

$$\left. +224\,x^2\sinh\frac{3x}{2} + 12\,x^4\sinh\frac{3x}{2} - 768\,\sinh\frac{5x}{2} - 64\,x^2\sinh\frac{5x}{2}\right\}$$

$$(21)$$

obtain the graphical expansion for $W_3(x_0)$ (the multiplicities of the diagrams have been canceled by the $n!$-denominators of the couplings $g_n/\hbar n!$; now the vertices represent only $g_n/\hbar$):

$$\beta W_3(x_0) = V(x_0) - \frac{1}{2}\bigcirc + \left(\frac{1}{2}\bigcirc + \frac{1}{8}\infty\right)$$

$$-\frac{1}{2}\left[\frac{1}{2}\bigcirc + \frac{1}{2}\infty + \frac{1}{6}\ominus + \frac{1}{24}\ominus\!\!\!- + \frac{1}{8}\infty\!\!\!\bigcirc\right]$$

$$+\frac{1}{6}\left[\bigcirc + \frac{3}{2}\ominus + 3\left(\frac{1}{4}\infty + \frac{1}{2}\infty\right)\right.$$

$$+3\left(\frac{1}{4}\ \raisebox{-3pt}{\includegraphics[height=20pt]{fig1}}\ +\frac{1}{4}\ \raisebox{-3pt}{\includegraphics[height=20pt]{fig2}}\right)$$

$$+3\left(\frac{1}{4}\ \raisebox{-3pt}{\includegraphics[height=20pt]{fig3}}\ +\frac{1}{6}\ \raisebox{-3pt}{\includegraphics[height=20pt]{fig4}}\ +\frac{1}{4}\ \raisebox{-3pt}{\includegraphics[height=20pt]{fig5}}\right)$$

$$+\left(\frac{3}{16}\ \raisebox{-3pt}{\includegraphics[height=20pt]{fig6}}\ +\frac{1}{4}\ \raisebox{-3pt}{\includegraphics[height=20pt]{fig7}}\ +\frac{1}{8}\ \raisebox{-3pt}{\includegraphics[height=20pt]{fig8}}\ +\frac{1}{8}\ \raisebox{-3pt}{\includegraphics[height=20pt]{fig9}}\right)\Big].$$

corresponding to the analytic expression (in the same order)

$$\begin{aligned}
W_3 &= V(x_0) + V_\Omega^{x_0} + \left(\frac{g_2}{2}a^2 + \frac{g_4}{8}a^4\right) \\
&\quad - \frac{1}{2!\hbar\Omega}\left[\frac{g_2^2}{2}a_2^4 + \frac{g_2 g_4}{2}a_2^4 a^2 + \frac{g_3^2}{6}a_2^6 + \frac{g_4^2}{24}a_2^8 + \frac{g_4^2}{8}a_2^4 a^4\right] \\
&\quad + \frac{1}{3!\hbar^2\Omega^2}\bigg[g_2^3 a_3^6 + 3\frac{g_2 g_3^2}{2}a_3^8 + 3\left(\frac{g_2^2 g_4}{4}(a_2^4)^2 + \frac{g_2^2 g_4}{2}a_3^6 a^2\right) \\
&\qquad + 3\left(\frac{g_3^2 g_4}{4}a_3^8 a^2 + \frac{g_3^2 g_4}{4}a_3^{10}\right) \\
&\qquad + 3\left(\frac{g_2 g_4^2}{4}(a_2^4)^2 a^2 + \frac{g_2 g_4^2}{6}a_{3'}^{10} + \frac{g_2 g_4^2}{4}a_3^6 a^4\right) \\
&\qquad + \left(\frac{3g_4^3}{16}(a_2^4 a^2)^2 + \frac{g_4^3}{4}a_{3'}^{10}a^2 + \frac{g_4^3}{8}a_3^{12} + \frac{g_4^3}{8}a_3^6 a^4\right)\bigg]
\end{aligned} \qquad (22)$$

The best $\Omega(x_0)$ is found numerically by searching for the roots of the first derivate of $W_3$ with respect to $\Omega$. There are many solutions and we must pick the right one, which we take to be the one closest to the unique solution found at the lowest order (where the Jensen-Peierls inequality ensured its existence).

Fig. 1 shows the approximations $W_{1,2,3}$ over the variational parameter $\Omega$ for $g = 4$ and $\beta = 1$ at $x_0 = 0$. The new extremum lies indeed near the old one. Note that $W_2$ has no extremum as remarked before, but the curve has a very small slope and the point of smallest $\Omega$-dependence is easily identified. In fact, the $\Omega$-dependence decreases rapidly with increasing order reflecting the fact that the exact $W_\infty$ is completely independent of the choice of $\Omega$.

The corresponding results for the free energy

$$F = -\frac{1}{\beta}\ln Z$$

are listed in Tab. 1 and compared with the precise numbers $F_{\text{ex}}$ of Ref. [7].

To third order, the new approximation gives energies which are better than the lowest order results by a factor of 30 to 50. They differ from the exact results only in the forth digit.

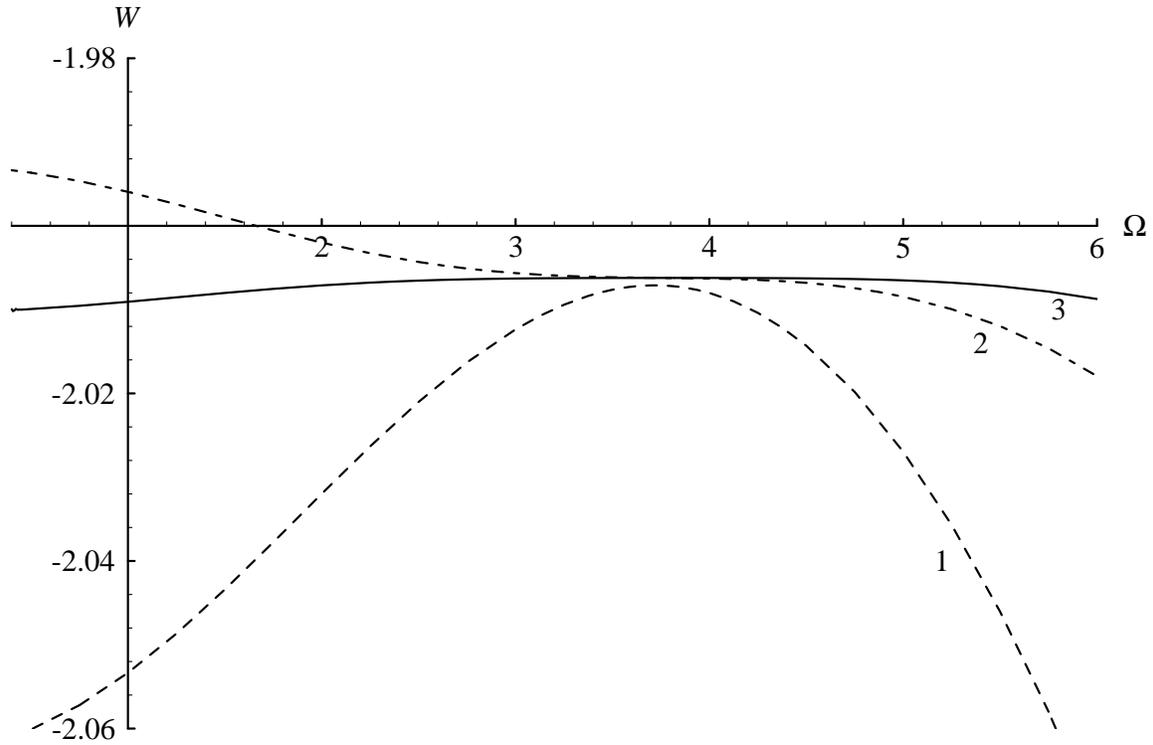

Figure 1: Effective potential $V_{\text{eff}}$ versus $\Omega$ for $g = 4$ and $\beta = 1$ at $x_0 = 0$. The curves 1,2,3 show the approximations $W_{1,2,3}$ respectively. While $W_1$ and $W_3$ have an extremum at $\Omega \approx 3,7$, the second-order approximation $W_2$ has only extrema in the complex plane.

| $g$ | $\beta$ | $F_1$ | $F_3$ | $F_{\text{ex}}$ |
|---:|---:|---|---|---|
| 0.002 | 2.0 | 0.427937 | 0.427937 | 0.427741 |
| 0.4 | 1.0 | 0.226084 | 0.226075 | 0.226074 |
| | 5.0 | 0.559155 | 0.558678 | 0.558675 |
| 2.0 | 1.0 | 0.492685 | 0.492578 | 0.492579 |
| | 5.0 | 0.699431 | 0.696180 | 0.696118 |
| | 10.0 | 0.700934 | 0.696285 | 0.696176 |
| 4.0 | 1.0 | 0.657396 | 0.6571051 | 0.6571049 |
| | 5.0 | 0.809835 | 0.803911 | 0.803758 |
| 20 | 1.0 | 1.18102 | 1.17864 | 1.17863 |
| | 5.0 | 1.24158 | 1.22516 | 1.22459 |
| | 10.0 | 1.24353 | 1.22515 | 1.22459 |
| 200 | 5.0 | 2.54587 | 2.50117 | 2.49971 |
| 2000 | 0.1 | 2.6997 | 2.69834 | 2.69834 |
| | 1.0 | 5.40827 | 5.32319 | 5.31989 |
| | 10.0 | 5.4525 | 5.3225 | 5.3199 |
| 80000 | 0.1 | 18.1517 | 18.0470 | 18.0451 |
| | 3.0 | 18.501 | 18.146 | 18.137 |

Table 1: Free energy of the anharmonic oscillators with the potential $V(x) = 1x^2/2 + gx^4/4$ for different $g$ and $\beta = 1/kT$.

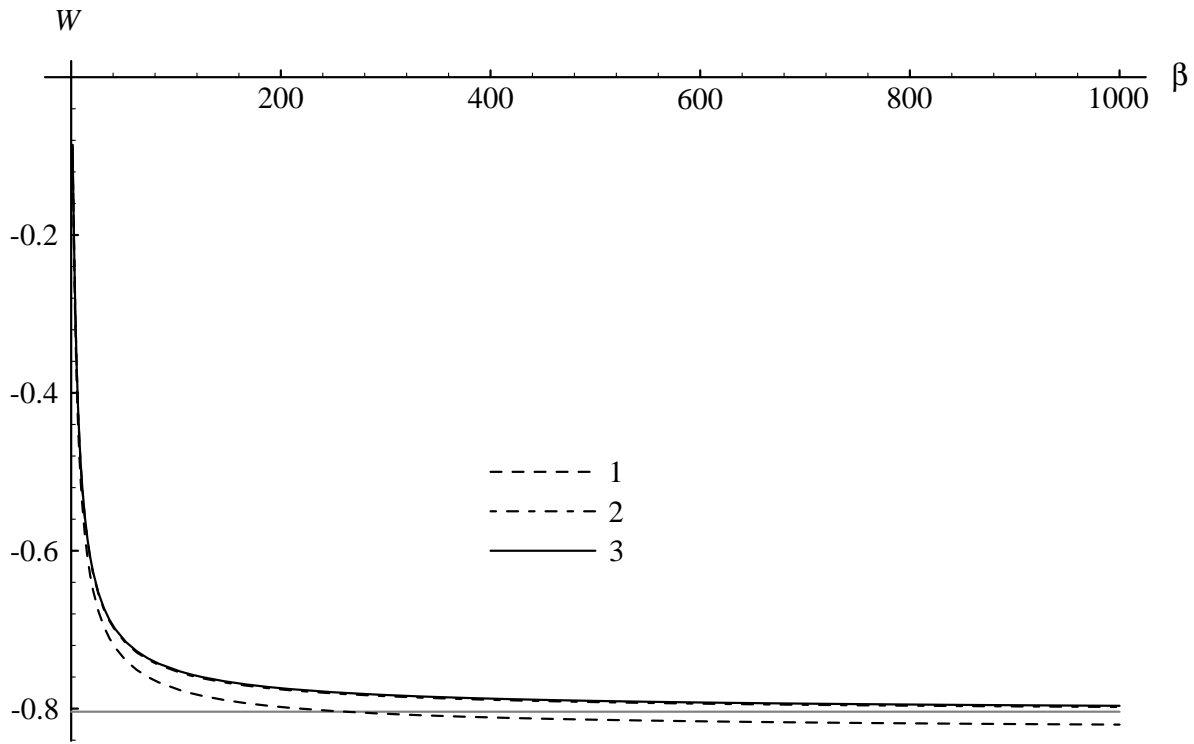

Figure 2: The approximations $W_{1,2,3}$ for the effective classical potential $V_{\text{eff}}$ versus $\beta$ for $g = 4$ and $x_0 = 0$ with optimized $\Omega$. The curves 1,2,3 shows the approximations $W_{1,2,3}$ respectively. The dotted line is the asymptote of $W_3$.

In the high-temperature limit, all $W_N$ tend to the classical results (which, in turn, becomes exact). It is therefore not astonishing that for small $\beta$, the approximations $W_3$ and $W_1$ are practically indistinguishable. Fig. 2 shows the dependence of $W_{1,2,3}$ on $\beta$. For increasing $\beta$, the difference between the first approximation and the other curves increases.

The worst possible case is $T = 0$, where the diagrams become most simple and have been given before [6].

# 1 Conclusion

We have presented a systematic improvement of the Feynman-Kleinert approximation up to third-order. For the anharmonic oscillator the free energy was obtained with an error less than 0.04 % for all coupling constants $g$ and all temperatures. In comparison with the previously available lowest-order approximation, the accuracy is increased by about a factor 30.

The virtue of the improved variational approach lies in the ability to yield very precise results for *all* coupling constants and temperatures. This is in contrast to other

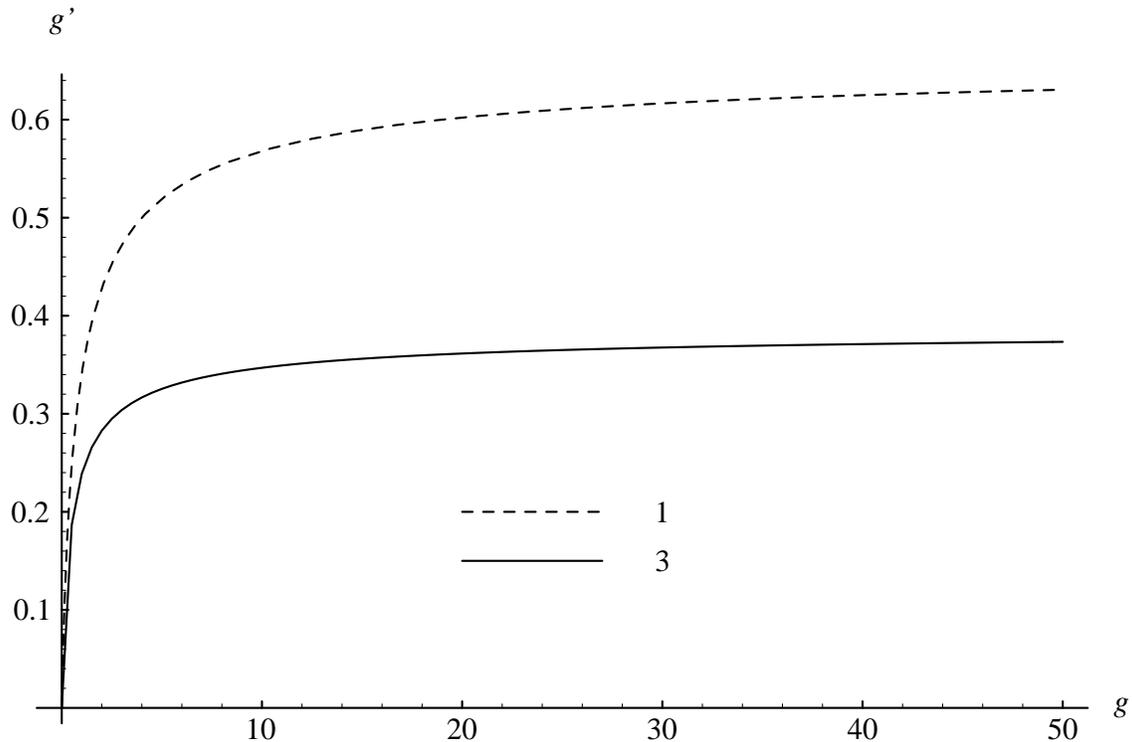

Figure 3: The effective dimensionless coupling constant $g'\hbar/\Omega^3 = g\hbar/\Omega^3(g)$ of the new perturbation series versus $g\hbar/\Omega^3$ in the first and third approximation $W_1$ and $W_3$, respectivly (axes are labeled setting $\hbar = \Omega = 1$).

methods which are applicable only in selected limiting temperature or coupling constant regimes.

The improvement with respect to the ordinary perturbation expansion is due to the fact that the new perturbative expansion has an effective coupling constant $g'\hbar/\Omega^3 \equiv g\hbar/\Omega^3(g)$, which remains small even for $g \to \infty$ (see Figure 3).

Certainly, the accuracy can be increased by calculating higher-order terms. The complexity, however, increases rapidly with increasing orders. Thus higher-order calculations will require much more work.

In this note, we have presented only the free energy of the anharmonic oscillator. Other quantities like particle distributions can be obtained quite similarly. This will be done in future work.